\documentclass[12pt]{iopart}
\newcommand{\msr}{$\mu$SR}
\newcommand{\lnfeasof}{Ln\-Fe\-As\-O$_{1-x}$\-F$_{x}$}

\newcommand{\bm}{\boldsymbol}
\usepackage[pdftex]{graphicx}
\usepackage{iopams}
\usepackage{color}
\definecolor{red}{rgb}{0.85,.1,0}
\begin{document}

\title[Pressure effects in LnFeAsO]{Effect of external pressure on the magnetic properties of LnFeAsO (Ln = La, Ce, Pr, Sm)}

\author{Roberto De Renzi, Pietro Bonf\`a, and Marcello Mazzani}
\address{Dipartimento di Fisica and Unit\`a CNISM di Parma, I-43124 Parma, Italy}
\ead{roberto.derenzi@unipr.it}
\author{Samuele Sanna, Giacomo Prando\footnote{now at Leibnitz Institute IFW, Helmholtzstrasse 20, D-01069 Dresden, Germany}, and Pietro Carretta}
\address{Dipartimento di Fisica ``A.\ Volta'' and Unit\`a CNISM di Pavia, I-27100 Pavia, Italy}
\author{Rustem Khasanov, Alex Amato, and Hubertus Luetkens}
\address{Paul Scherrer Institut, CH-5232 Villigen PSI, Switzerland}
\author{Markus Bendele}
\address{University of Zurich, Winterthurer Strasse 190, 8057 Zurich}
\author{Fabio Bernardini and Sandro Massidda}
\address{CNR-SLACS and Universit\`a di Cagliari, I-09042 Monserrato, Italy}
\author{Andrea Palenzona, Matteo Tropeano  and Maurizio Vignolo}
\address{CNR-SPIN and Universit\`a di Genova, via Dodecaneso 33, I-16146 Genova, Italy}

\begin{abstract}
We investigate the effect of external pressure on magnetic order in undoped LnFeAsO (Ln = La, Ce, Pr, La) by using muon-spin relaxation measurements and ab-initio calculations. Both magnetic transition temperature $T_m$ and Fe magnetic moment decrease with external pressure. The effect is observed to be lanthanide dependent with the strongest response for Ln = La and the weakest for Ln = Sm. The trend is qualitatively in agreement with our DFT calculations. The same calculations allow us to assign a value of 0.68(2) $\mu_B$ to the Fe moment, obtained from an accurate determination of the muon sites. Our data further show that the magnetic lanthanide order transitions do not follow the simple trend of Fe, possibly as a consequence of the different $f$-electron overlap.
\end{abstract}

\maketitle

\section{\label{sec:intro}Introduction}
Superconductivity in iron based (IB) pnictides and chalcogenides originates from a magnetic parent, as in the cuprates. This remarkable analogy hides the well known fact that IB materials are always metallic, pointing to a different role of electron correlations in the two cases. Another noteworthy difference is that the phase diagram of the IB compounds, from the magnetic to the superconducting phase, can be spanned not only by doping, as in the case of cuprates, but also by pressure \cite{Takahashi2008, Lorenz2008, Okada2008, Kawakami2009, Kotegawa2009, Chu2009, Athena2011, Khasanov2011} and by nominally isovalent substitutions \cite{Tropeano2009, Maroni2011, Sanna2011}.

The \lnfeasof\ family where Ln is a lanthanide, also referred to as 1111, is particularly relevant because it still holds the record superconducting critical temperature $T_c=56$K \cite{Wu2009}. The magnetic and superconducting properties of 1111 are however the least sensitive to pressure \cite{Chu2009, Athena2011, Khasanov2011,Lu2008} among IB materials, and the pressure dependence of the magnetic moment in the parent members is not known. This motivates the present investigation on the effect of pressure on the magnetic ordering temperature and on the magnitude of the zero temperature ordered moment for Ln = La, Ce, Pr, Sm.

These four compounds have already been investigated at ambient pressure by neutron scattering, providing the ordering vector of the spin density wave (SDW), [1/2 1/2 1/2] and [1/2 1/2 0] in the tetragonal setting, for La and for Pr, Ce, respectively \cite{Lumsden2010}. The value of the magnetic moment on Fe was also obtained and apparently it varies between 0.25 $\mu_B$ and 0.8 $\mu_B$, but the influence of the rare earth moment on the measurements and some earlier results may be questioned.  Ambient pressure muon spin rotation \msr\  \cite{Luetkens2009,Sanna2009,Maeter2009,Shiroka2011} and Moessbauer  measurements \cite{McGuire2009} indicate instead that the magnetic moment on iron has very similar values in all compounds (within 0.06\% of each other). At low temperature ($T<5$ K) muons detect also the ordering of the rare earths. In the special case of Ce the polarization of the rare earth by the ordered Fe lattice, indicating a strong Fe-Ce hybridization, provides an additional contribution to the local field at the muon site, observed already well above the magnetic ordering of the Ce sublattice. \cite{Maeter2009}

In this work we describe the muon results under pressure, from which we extract the pressure and the temperature dependence of the Fe magnetic moment, on the basis of an ab-initio determination of the muon site. The same ab-initio calculations provide insight in the discussion of the Fe moment behaviour.

\section{\label{sec:methods} Methods}

In pressure experiments a large fraction of the muons, roughly 50\%, stops in the pressure cell (described elsewhere in details \cite{Khasanov2011,Khasanov2012}), providing a background contribution that has to be separated from the sample signal in the data analysis.

In these conditions the most straightforward method to determine the magnetic transition temperature under pressure is by means of \msr\ measurements in a weak external magnetic field $\bm{H}$, applied orthogonal to the initial muon spin direction, in the so called weak transverse field (WTF) configuration. In this case the contribution  to the asymmetry from muons experiencing a vanishing internal spontaneous magnetisation can be accurately determined.  Muons stopping in a non magnetic environment produce long lived oscillations, which reflects the coherent muon precession around the external field $\bm{B}_0=\mu_0 \bm{H}$, whereas muons stopping in magnetically ordered parts of the sample give rise to a more complex, distinguishable signal, reflecting the vector combination $\bm{B}_\mu=\bm{B}_i+\bm{B}_0$ of the internal and applied fields. The random orientation of the grains in a polycrystalline sample leads to a broad distribution of precession frequencies.

A distinct experiment can be performed in zero applied field (ZF). In this case muons precess around the internal field alone, which may be then more accurately extracted from the oscillating muon asymmetry, at the price of a less straightforward identification of the transition temperature, where the precessing component disappears. Ambient pressure ZF \msr\ experiments can be further calibrated in separate experiments in the absence of the pressure cell, stopping therefore nearly all the muons in the sample.

In the following we describe in more details the experimental results for a couple of our LnFeAsO samples, namely Ln=La, Ce, and we conclude with a summary of the results for the entire series of samples

\subsection{\label{subsec:WTFmuSR} Determination of magnetic transition $T_m$}

\begin{figure}
\begin{center}
\includegraphics[width=90mm]{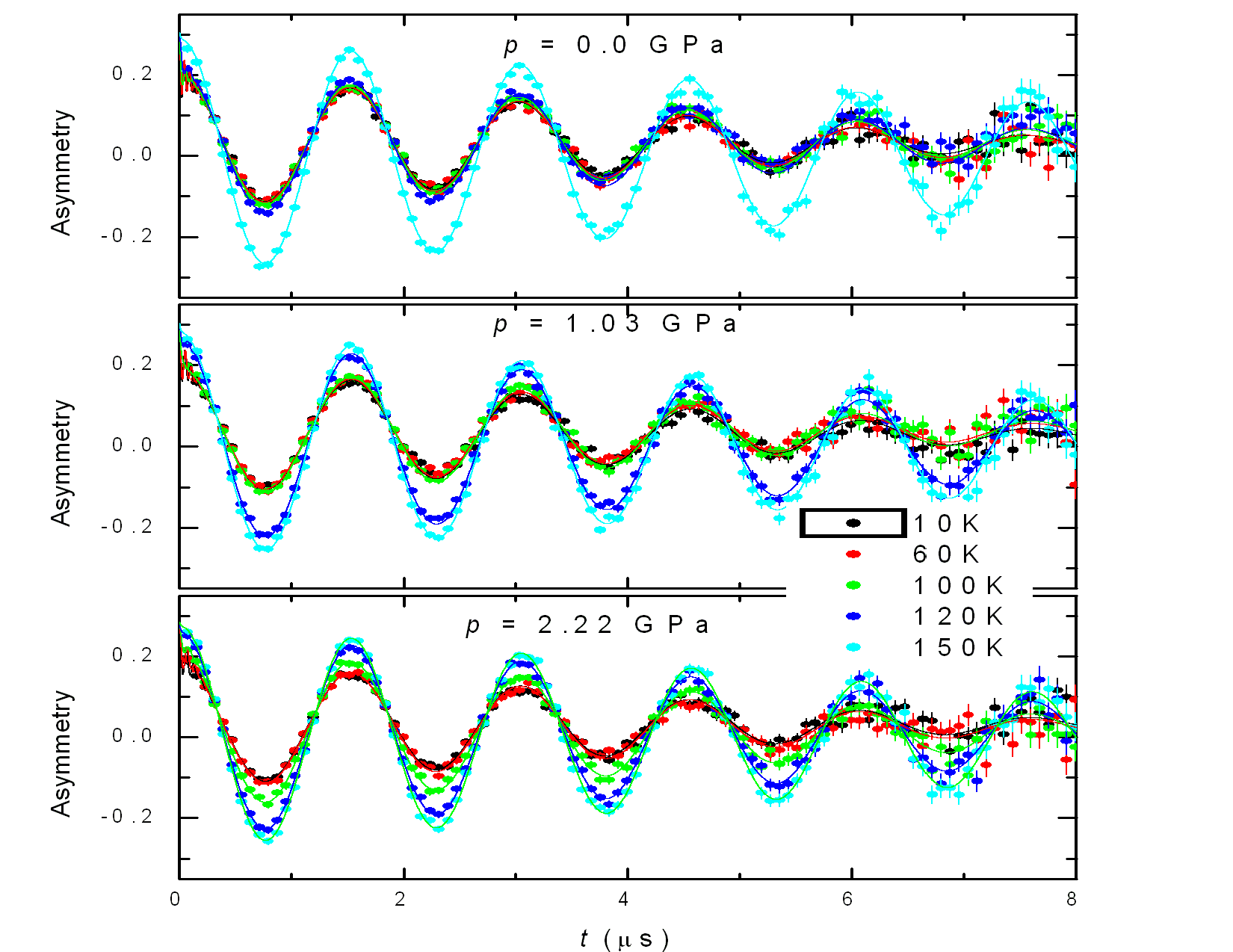}%
\caption{\label{fig:WTFmusr}(Colour on-line)  Typical time evolution of the WTF-$\mu$SR asymmetry for the LaFeAsO sample with an external field of $B_0=5$ mT, applied perpendicular to the initial muon polarisation. Above the magnetic transition both the pressure cell and the sample contribute to the initial oscillating amplitude, whereas only the pressure cell contributes at low temperatures.}
\end{center}
\end{figure}

\begin{figure}
\begin{center}
\includegraphics[width=90mm]{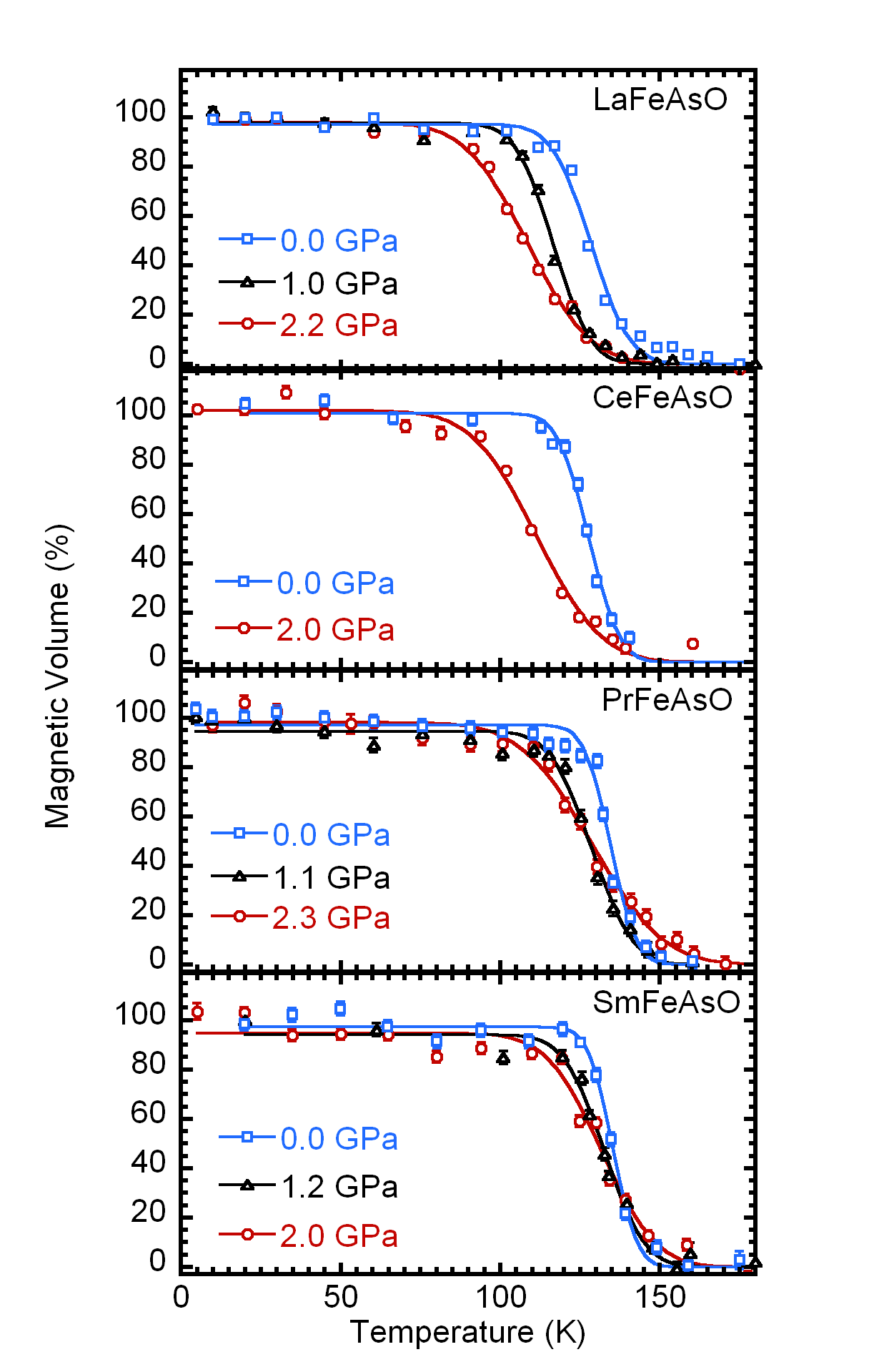}%
\caption{\label{fig:MagVol}(Colour on-line) Temperature dependence of the magnetic volume fraction at different external pressure for the LnFeAsO samples with Ln=La, Ce, Pr, Sm.}
\end{center}
\end{figure}

Let us first describe the WTF procedure in the representative case of the Ln=La sample.
We performed WTF \msr\ measurements between 5 K and 200 K in an external field of $\mu_0H= 5$ mT, perpendicular to the initial muon spin direction $\bm{S}_\mu$.  The three panels of Fig.~\ref{fig:WTFmusr} display the time evolution of the WTF-\msr\ asymmetry at different temperatures, for three fixed applied pressure values. The solid curves in the figure represent the global fit at each pressure to the following function:

\begin{eqnarray}
\label{eq:TFasymmetry}
\lefteqn{
{{\cal A}_{\mathrm{WTF}}(t)} = \left( a_{PC}  e^{-\sigma_{PC}^2 t^2/2}  + a_{PM} \, e^{-\lambda_{PM} t}\right) \cos(\gamma B_0 t) } \nonumber\\ & &
+ a_T e^{-\sigma^2_T t^2/2} \cos(\gamma B_i t) + a_L \, e^{-\lambda_L t}
\end{eqnarray}
where  $\gamma/2\pi= 135.5$ MHz/T is the muon gyromagnetic ratio. The first term is the contribution of muons stopping in the pressure cell (cell asymmetry), and its small relaxation rate, $\sigma_{PC}=0.5\ \mu s^{-1}$, has been calibrated separately with the empty cell. The initial asymmetry $a_{PC}$ is fitted globally, imposing a common value for all temperatures. The second term is due to muons implanted in the part of the sample which is still in the paramagnetic phase (paramagnetic asymmetry $a_{PM}$). The global procedure converges rapidly thanks to the fact that the paramagnetic asymmetry vanishes for $T\ll T_m$, as it was verified for all the samples at ambient pressure in a separate WTF experiment without the pressure cell.  Muons stopping in a local environment that is magnetically ordered are described by the last two terms (the magnetic asymmetry), which reflect respectively the transverse ($\bm{B}_\mu\perp \bm{S}_\mu$) and the longitudinal ($\bm{B}_\mu \parallel \bm{S}_\mu$) components of the total local field, averaged over all grain orientations.

\begin{figure}
\begin{center}
\includegraphics[width=90mm]{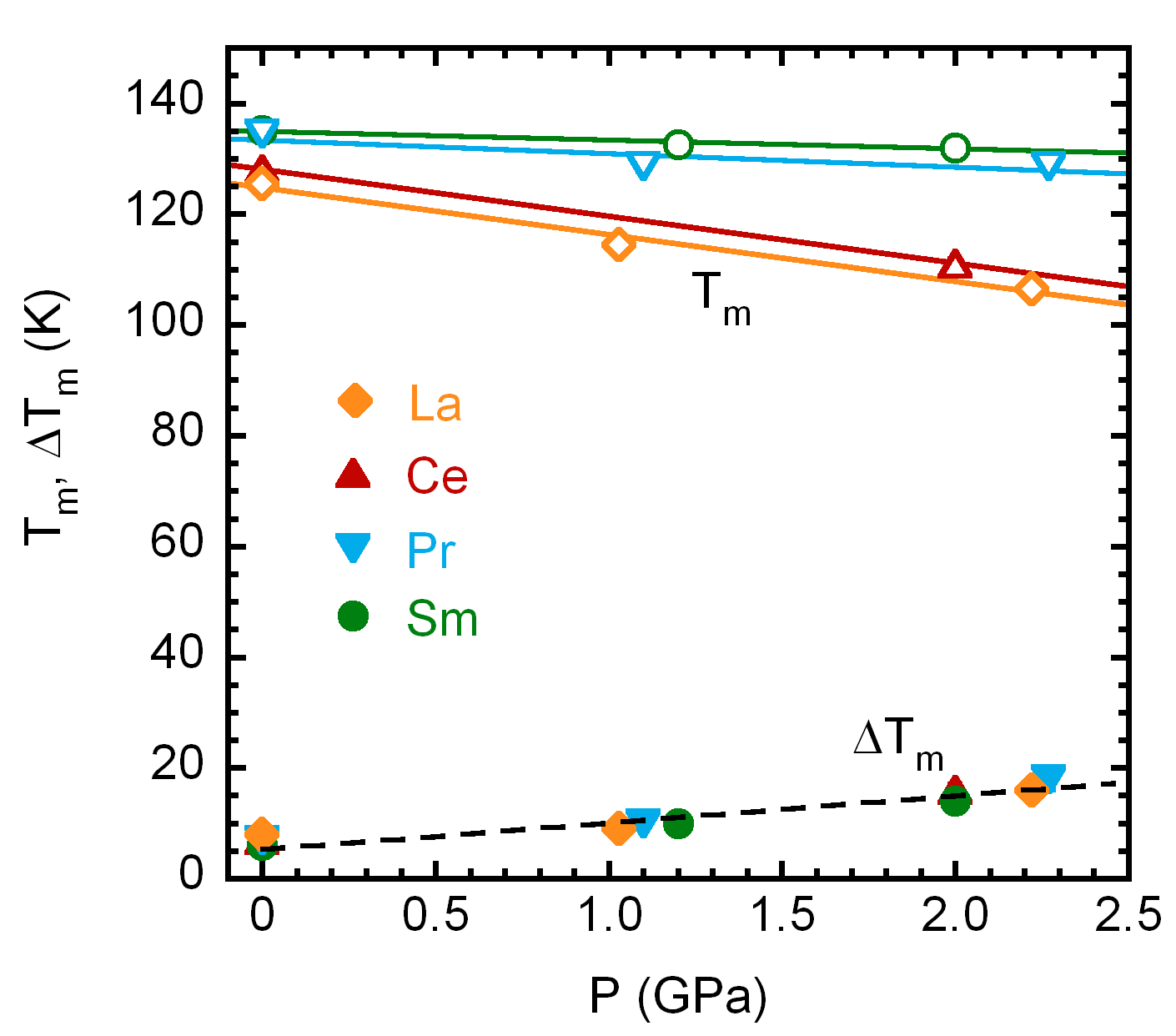}%
\caption{\label{fig:TmP}(Colour on-line) Pressure dependence of the magnetic transition temperatures $T_m$ and transition widths $\Delta T_m$ as a function of external pressure for the LnFeAsO samples with Ln=La, Ce, Pr, Sm}
\end{center}
\end{figure}

 The temperature dependent magnetic fraction of the sample is evaluated as $1-a_{PM}/a_0$ where $a_0$ is the total muon asymmetry from the sample, obtained from the global fit at high temperatures, well above the magnetic transition $T_m$. The same procedure has been applied to all our samples. The values of the magnetic volume fraction for LnFeAsO Ln=La, Ce, Pr, Sm are displayed as a function of temperature in Fig.~\ref{fig:MagVol} for the different values of the applied pressure. The magnetic transitions and their widths have been evaluated by the best fit of each experimental data set to the function $\textrm{erf}[(T-T_m)/(\sqrt{2}\Delta T_m))]$, which accounts for a Gaussian distribution of $T_m$.  These parameters are displayed in Fig.~\ref{fig:TmP}, together with their linear regressions (solid lines for $T_m$). The dashed line for $\Delta T_m$ is the pressure dependence of the width of the transition and it is indicative perhaps of the increasing inhomogeneity of the pressure within the cell.

\subsection{\label{subsec:ZFmuSR} Determination of the internal field at the muon sites $B_i$}

\begin{figure}
\begin{center}
\includegraphics[width=90mm]{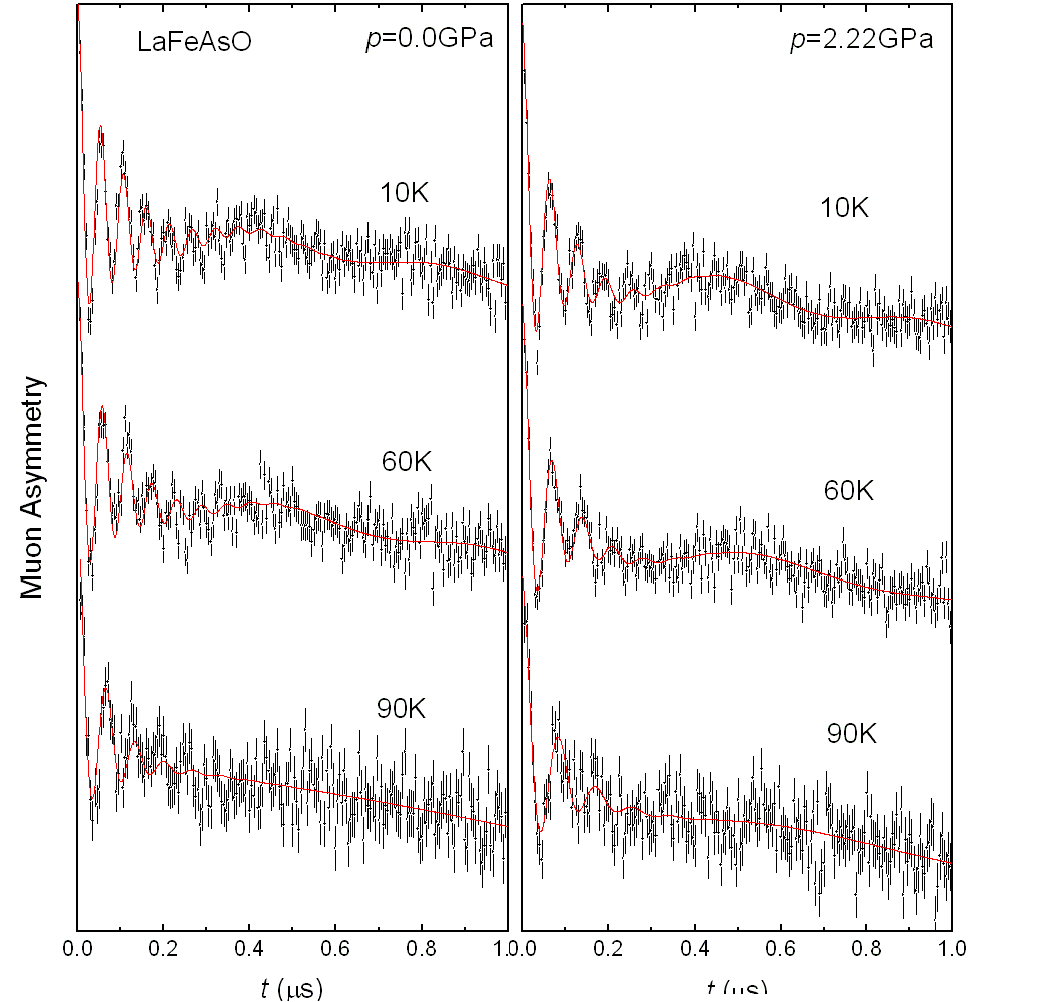}%
\caption{\label{fig:ZFmusrLa}(Colour on-line) Time evolution of the ZF-$\mu$SR asymmetry for the LaFeAsO sample at ambient (left) and maximum (right) pressure, at three representative temperatures.}
\end{center}
\end{figure}

\begin{figure}
\begin{center}
\includegraphics[width=90mm]{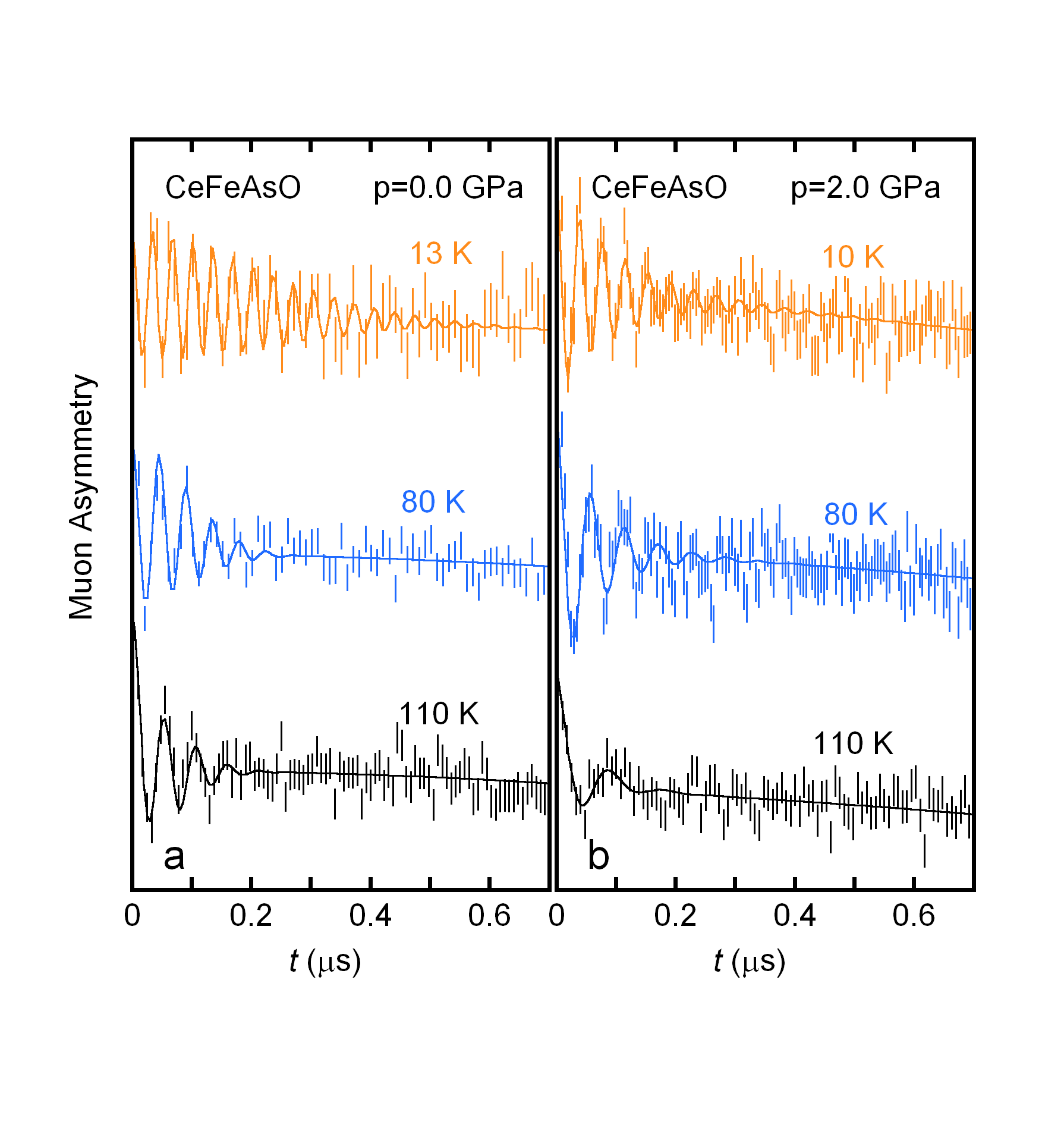}%
\caption{\label{fig:ZFmusrCe}(Colour on-line) Time evolution of the ZF-$\mu$SR asymmetry for the CeFeAsO sample at ambient (left) and maximum (right) pressure, at three representative temperatures.}
\end{center}
\end{figure}

The most accurate determination of the internal fields and how they change as a function of temperature and pressure is obtained by separate ZF \msr\ experiments. The spontaneous internal fields $B_i$ at the muon site are proportional to the intensity of the staggered magnetic moment on iron, $m$, with an additional  contribution from magnetic rare earths below their ordering temperatures. Representative ZF-\msr\ spectra are shown in Fig.~\ref{fig:ZFmusrLa} and Fig.~\ref{fig:ZFmusrCe}, for Ln=La and Ce respectively. Equation \ref{eq:TFasymmetry} may be rewritten now for $B_0=0$ as:

\begin{eqnarray}
\label{eq:ZFasymmetry}
\lefteqn{
{{\cal A}_{\mathrm{ZF}}(t)} = a_{PC} \, e^{-\sigma_{PC}^2 t^2/2} + a_{PM} \, e^{-\lambda_{PM} t}  }  \nonumber\\ & & + \sum_{i=1,2} a_{Ti} \, e^{-\sigma^2_{Ti} t^2/2} \cos(\gamma B_i t) + a_L \, e^{-\lambda_L t}
\end{eqnarray}
where higher statistics ZF experiment allows us to recognise two distinct transverse components in the magnetic asymmetry, indicated by subscripts $i=1,2$. They are due to the two known muon stopping sites in 1111 compounds \cite{Maeter2009}, with $B_1$ corresponding to a site within the FeAs layer and $B_2$ corresponding to a site close to O ions, for which we find an occupancy ratio of $a_{T1}/a_{T2}\sim 4$. The corresponding longitudinal components cannot be resolved in the last term of the Eq.~\ref{eq:ZFasymmetry}.

\begin{figure}
\begin{center}
\includegraphics[width=90mm]{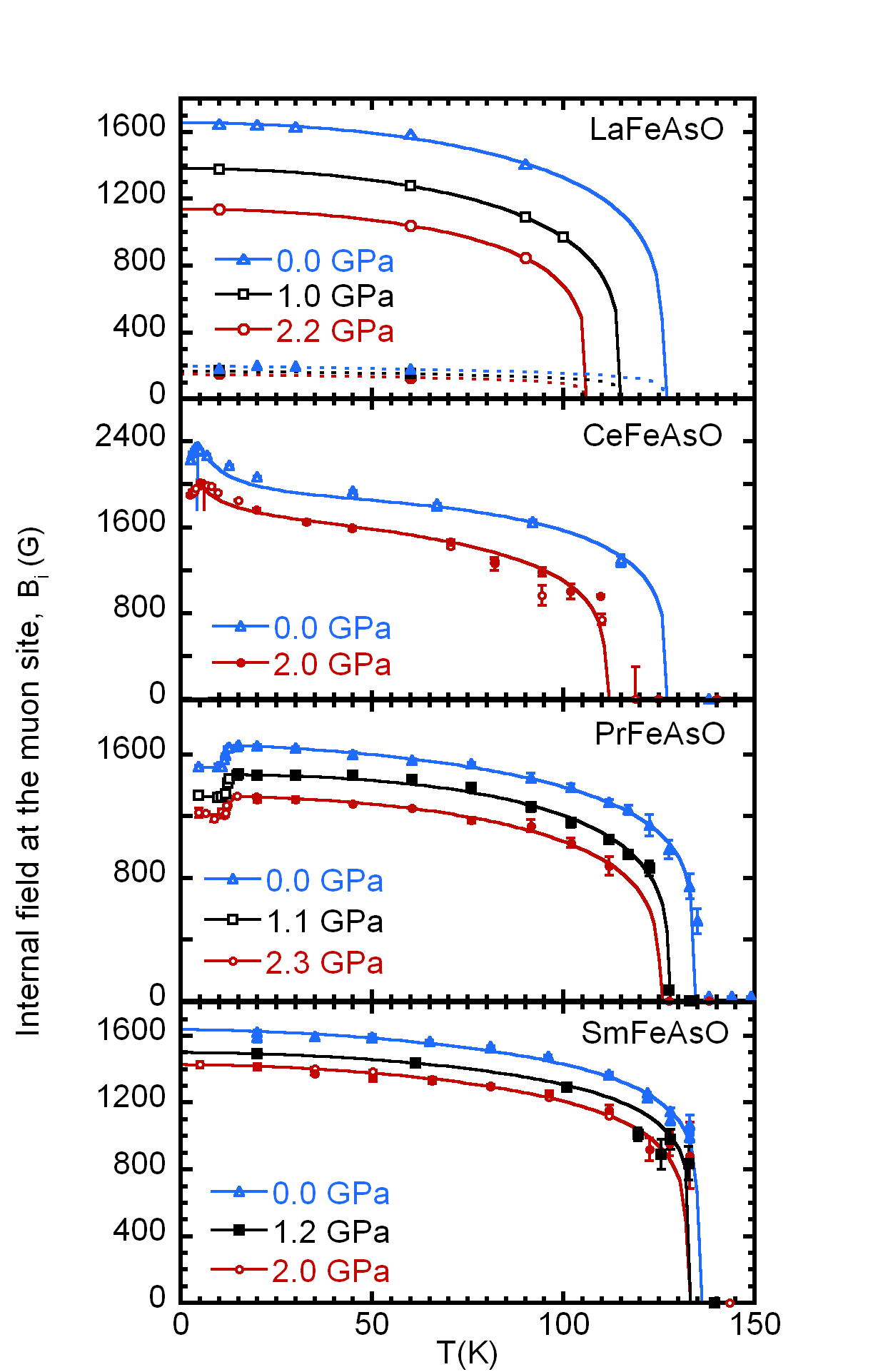}%
\caption{\label{fig:BmuT}(Colour on-line) Temperature dependence of the two internal fields, $B_i$, $i=1,2$: open symbols from ZF \msr, filled symbols from WTF \msr. Solid lines are the phenomenological fit of $B_1(T)$ to Eq.~\ref{eq:fitBmu} (Ln = La, Pr, Sm) and to \ref{eq:fitBmuCe} (Ln = Ce); the dashed line for Ln = La is the fit of the lower $B_2(T)$ field.}
\end{center}
\end{figure}

The low field $B_2$ is accurately detected only for the Ln=La sample and for $T<70$ K (Fig.~\ref{fig:ZFmusrLa}), reaching the value $B_2(0)\sim 18$ mT at ambient pressure, in agreement with Ref.~\cite{Klauss2008}. In the other investigated LnFeAsO compounds this low precession is over-damped due to the broader field distribution at this muon site. This is justified by the  stronger magnetic perturbation provided by the magnetic rare earth, much closer to this than to the other site. For the same reason the muon field $B_1$ can be easily detected almost up to the transition temperature. The two fields are displayed as a function of temperature in Fig.~\ref{fig:BmuT}.

\begin{figure}
\begin{center}
\includegraphics[width=90mm]{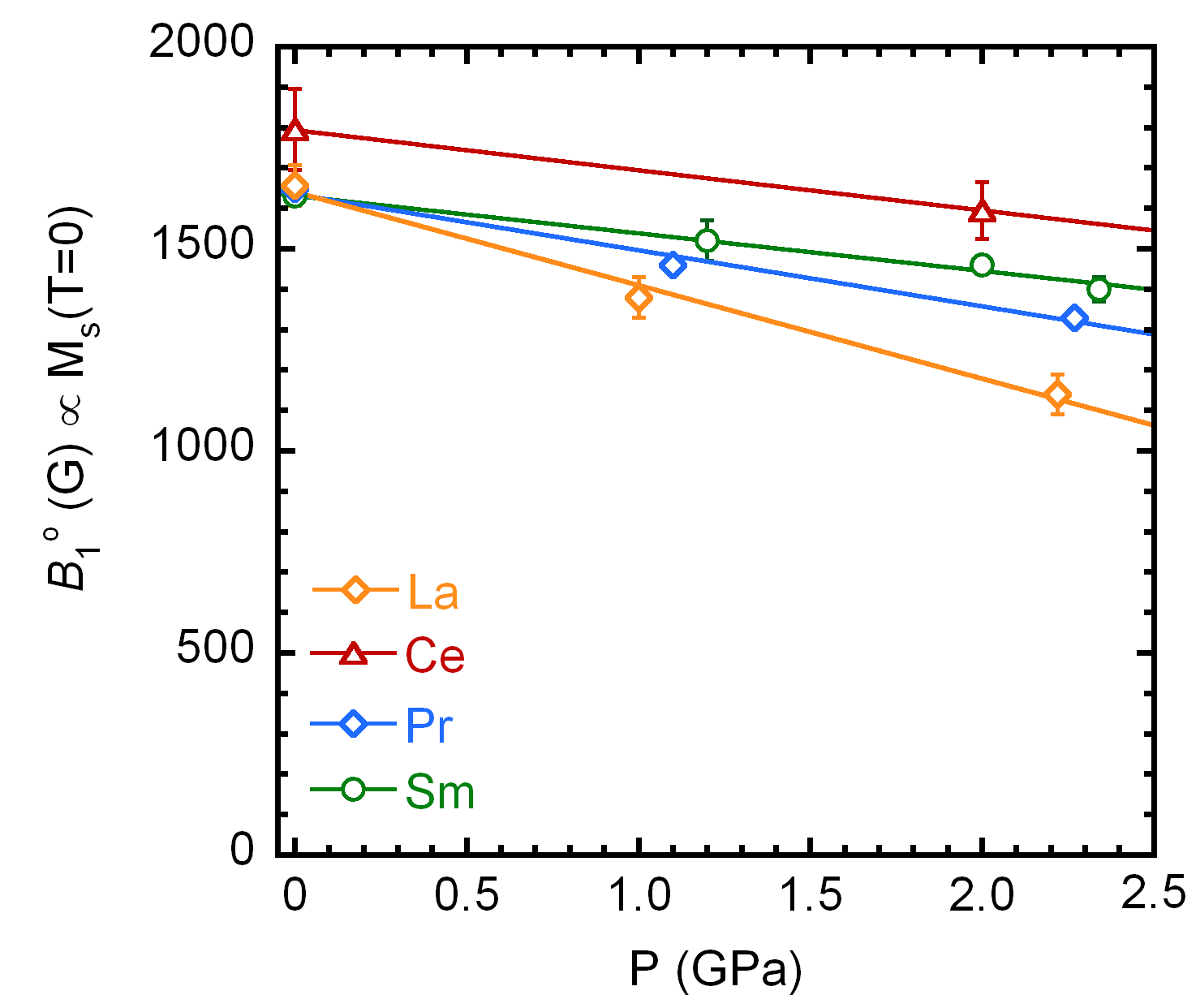}%
\caption{\label{fig:BmuP}(Colour on-line) Pressure dependence of the internal field at the most populated muon site, extrapolated at zero temperature, $B_1(0)$, for the LnFeAsO samples with Ln=La, Ce, Pr, Sm.
Straight lines are the best fit to a linear behaviour.}
\end{center}
\end{figure}

We are interested in particular in the effect of pressure on the zero temperature value of the staggered magnetic moment of Fe, $m\propto B_i(0)$. The extrapolation to zero temperature must avoid including any very low temperature variation of the internal field, due to the onset of rare-earth ordering. To this end the internal field has been fitted for $T>T_N^{Ln}$ to the phenomenological fit curve

\begin{eqnarray}
\label{eq:fitBmu}
B_i(T) &=& B_i(0)\left[1-(T/{T_m})^\alpha \right]^\beta,
\end{eqnarray}
that interpolates between a low temperature Bloch-like regime and the critical regime, close to $T_m$.

In the case Ln = Ce the molecular field of Fe on the paramagnetic Ce moments \cite{Maeter2009} produces and additional non negligible Curie-Weiss contribution from Ce at the muon:
\begin{eqnarray}
\label{eq:fitBmuCe}
B_i(T) &=& B_i(0)\left[1-\left(\frac T {T_m}\right)^\alpha \right]^\beta \left(1+ \frac C {T-T_N^{Ce}}\right)
\end{eqnarray}
The fits have been done by using the same $T_m$ values reported in Figure \ref{fig:TmP}. The power coefficients are $1.5\le\alpha\le2.5$ and $0.15\le\beta\le 0.2$ for all the samples.
The results of these fits are shown in Figure ~\ref{fig:BmuP} for $B_1(0)$ as a function of pressure for the all samples under investigation.

\subsection{\label{subsec:ZFmuSRLn} Determination of the magnetic transition of Ln=Sm,Pr,Ce sub-lattice, $T_N^{Ln}$}

\begin{figure}
\begin{center}
\includegraphics[width=90mm]{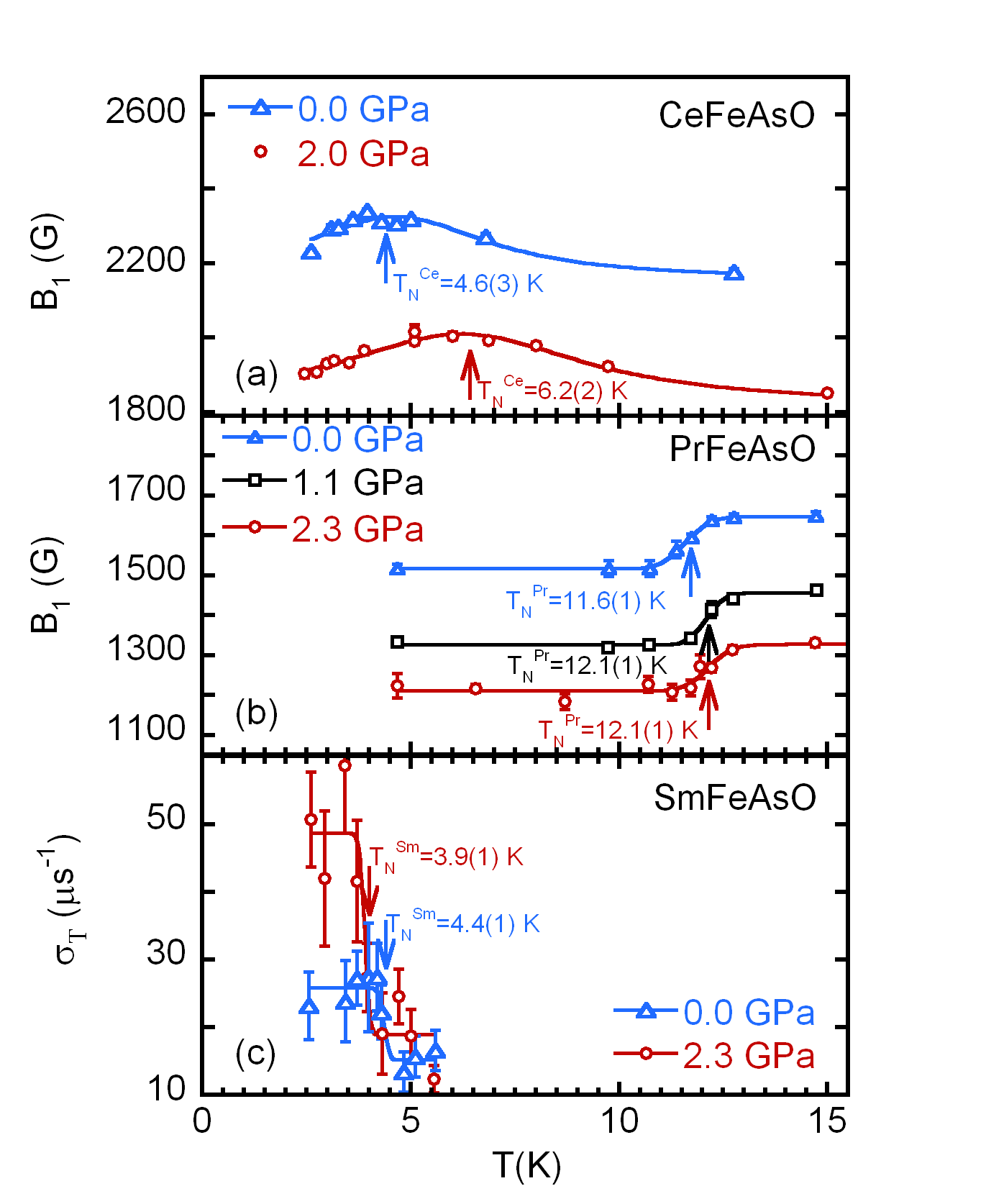}%
\caption{\label{fig:TNLn}(Colour on-line) Temperature dependence of the internal field,$B_\mu(T)$, for Ln = Ce (a) and Pr (b) and of the depolarisation rate of the transverse fraction, $\sigma_T(T)$, for Ln = Sm (c) nearby the antiferromagnetic transition of the Ln sub-lattice $T_N^{Ln}$}
\end{center}
\end{figure}

It is known that the at ambient pressure the magnetic Ln ions orders antiferromagnetically below $T_N= 4.6$, 4.4 and 11 K  for Ln=Sm, Pr and Ce respectively \cite{Maeter2009}. This magnetic transition introduces a Ln-dependent contribution to the muon precession frequency in the ZF-$\mu$SR signal, i.e. to the internal field $B_\mu$ of Eq.~\ref{eq:ZFasymmetry}. At ambient pressure the magnetic transitions of Ln = Pr, Ce are second-order-like \cite{Maeter2009} and produces  a reduction of $B_\mu(T)$ in the case of Pr, whereas for Ce the field shows a broad peak. Panel a and b of Figure \ref{fig:TNLn} show that the same behaviour is found also under external pressure. $T_N^{Ln}$ is found to increase for both rare earths, being $T_N^{Pr}=12.1(1)$ and $T_N^{Ce}=6.2(2)$ K at the maximum applied pressure $p=2.3$ and 2 GPa, respectively.

On the other hand, the magnetic transition of Sm in SmFeAsO at ambient pressure is discontinuous at $T_N^{Sm}$ (first-order) \cite{Maeter2009} and it is accompanied by a change of the magnetic unit cell. Below $T_N^{Sm}$ the single muon precession frequency line due to the order of Fe moments develops two smaller amplitude satellite lines. Unfortunately these satellites cannot be resolved in the presence of the large pressure cell background signal, and show up just as an increased static relaxation. The sharp change of the depolarisation rate of the transverse component, proportional to the width of the of the internal field distribution, is however sufficient to signal the onset of Sm order and its pressure dependence. Panel a in Fig.~\ref{fig:TNLn} shows that the maximum applied pressure $p=2.3$ GPa produces in this case a {\em reduction} of the magnetic transition to $T_N^{Sm}=3.8(1)$ K.

\section{\label{subsec:theoretical}Electronic structure and muon field calculations}
Electronic structure calculations are performed within the Density Functional Theory (DFT)
and local spin density approximation as parametrised by Perdew and Wang \cite{PeWa} plus
Hubbard corrections (LSDA+U) \cite{LDAU} as implemented in the all-electron LAPW code Wien2K \cite{wien2k}.
The muffin-tin radii for Ln = La, Ce, Sm, for Fe, As, and O are chosen equal to 2.3, 2.2, 2.0, and 1.9 Bohr respectively.  $R_{\rm MT} \times kmax=7$ is used as the plane-wave cutoff.
The $U$ on-site Coulomb parameter is set to 6 eV for Ce and Sm, and the collinear magnetic formalism is adopted.

The DFT calculation allows a straightforward identification of the muon site, corresponding in first approximation to the minimum of the electrostatic potential (i.e. the sum of the Hartree potential plus the local pseudopotential), reversed in sign, since the muon has a positive charge. In the case of the 1111 structure we identify three minima, the lowest two replicating those found in Ref.~\cite{Maeter2009} with a simple Thomas-Fermi approach (we quote here results for Ln=La).

\begin{figure}
\begin{center}
\includegraphics[width=90mm]{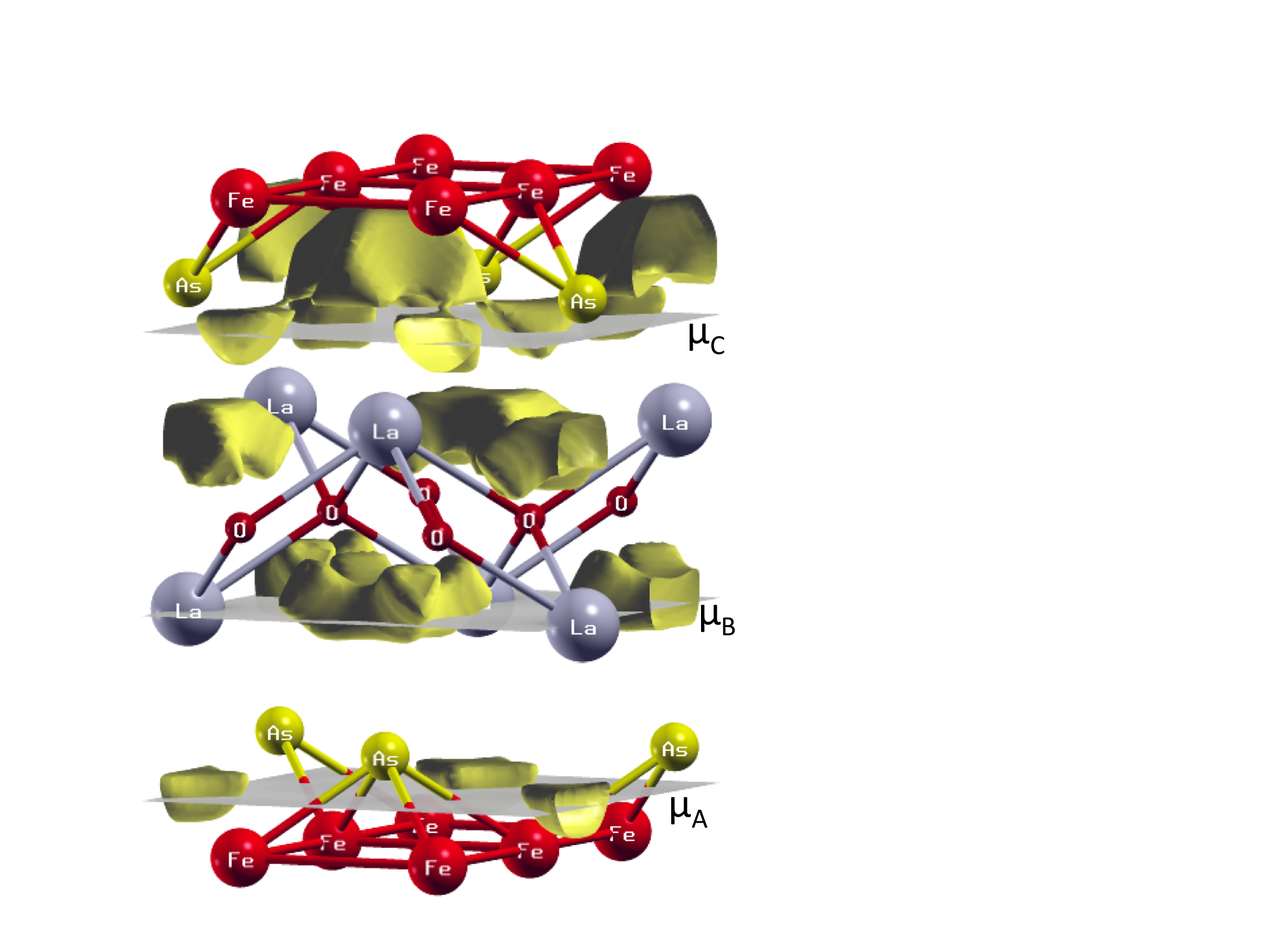}%
\caption{\label{fig:muon}(Colour on-line) Muons sites in the LaFeAsO unit cell by DFT. The golden light shaded areas represent the volume of zero-point displacement for the muon around the minimum.}
\end{center}
\end{figure}

Given the small muon mass ($m = 105.7$ MeV/c$^2$), we investigate possible effect of the zero-point-motion on the stability of the identified minima. We model the potential around each minimum as an anisotropic harmonic well
$V(r)= \frac{1}{2} m(\omega^2_xx^2+\omega^2_yy^2+\omega^2_zz^2)+V_0$ with eigenvalues given by
$E(n_x,n_y,n_z) = \hbar[\omega_x(n_x+1/2) +\omega_y(n_y+1/2) + \omega_z(n_z+1/2)] +V_0$.
The location, the minimum potential $V_0$, the zero point energy $E_0$ and the first excitation energy $E_1$ are
shown in Tab.~\ref{tab:muon}. The actual values of $E_1$ indicate that in every site only the ground state is occupied at room temperature. From these numbers we estimate the extent of the muon wave-function around each site by drawing the isosurface corresponding to $V(r) = V_0 + E_0$.
Figure \ref{fig:muon} shows these isosurfaces as a light-shaded areas. Assuming that in site $\alpha$ the muon overtakes any barrier lower or equal to $E^\alpha_0$, we see that site
A and B are disconnected, while site C and A form an interconnected network where the migration barrier is lower than $E^C_0$. Therefore site C is unstable and
we are lead to conclude that,
while the muon potential has three nonequivalent minima, only site A and B are surrounded by barriers high enough to guarantee muon confinement. These sites correspond roughly to those identified in Ref.~\cite{Maeter2009}.

\begin{table}[ht]
\begin{center}
\caption{Muon sites: cell coordinates $x, y, z$, potential $V$ (eV) referred to the minimum, zero point energy, first excited level $E_0, E_1$ (eV), and local dipolar field, assuming a value of the Fe moment  $\mu=0.68 \, \mu_B$ (see text).}
\label{tab:muon}
\begin{tabular}{llllllcc}
\hline
                site        & $x$  &  $y$  &  $z$ & $V-V_{\mathrm{A}}$ & $E_0$ & $E_1$ &  $B_\mu^d$ (mT)\\
\hline
$\mu_{\mathrm{A}}$    & 0.75 &  0.75  & 0.572 &       0     & 0.63 & 0.29 & 165 \\
$\mu_{\mathrm{B}}$    & 0.25 &  0.25  & 0.137 &     0.55    & 0.51 & 0.09 & 29\\
$\mu_{\mathrm{C}}$    & 0.5 &  0.0  & 0.703 &       0.94    & 0.54 & 0.25 & -\\
\hline
\end{tabular}
\end{center}
\end{table}

If we now calculate the  internal fields, $\bm{B}_i$ as the distant-dipole contribution from the Fe moments in the known \cite{Lumsden2010} magnetic structures, a Fe staggered magnetic moment $m=0.68(2)\,\mu_B$ along (110)$_T$ is required to obtain the experimental value $B_1(0)=0.165(5)$ T at site A. By the same token site B cannot be the main muon site, since it requires an unacceptably high Fe moment of 3.6 $\mu_B$ to reproduce $B_1(0)$. The field reported in Tab.~\ref{tab:muon} for site B, with 0.68 $\mu_B$ on Fe, agrees within a factor 1.6 with $B_2(0)$.

\begin{figure}[ht]
\begin{center}
\includegraphics[width=90mm]{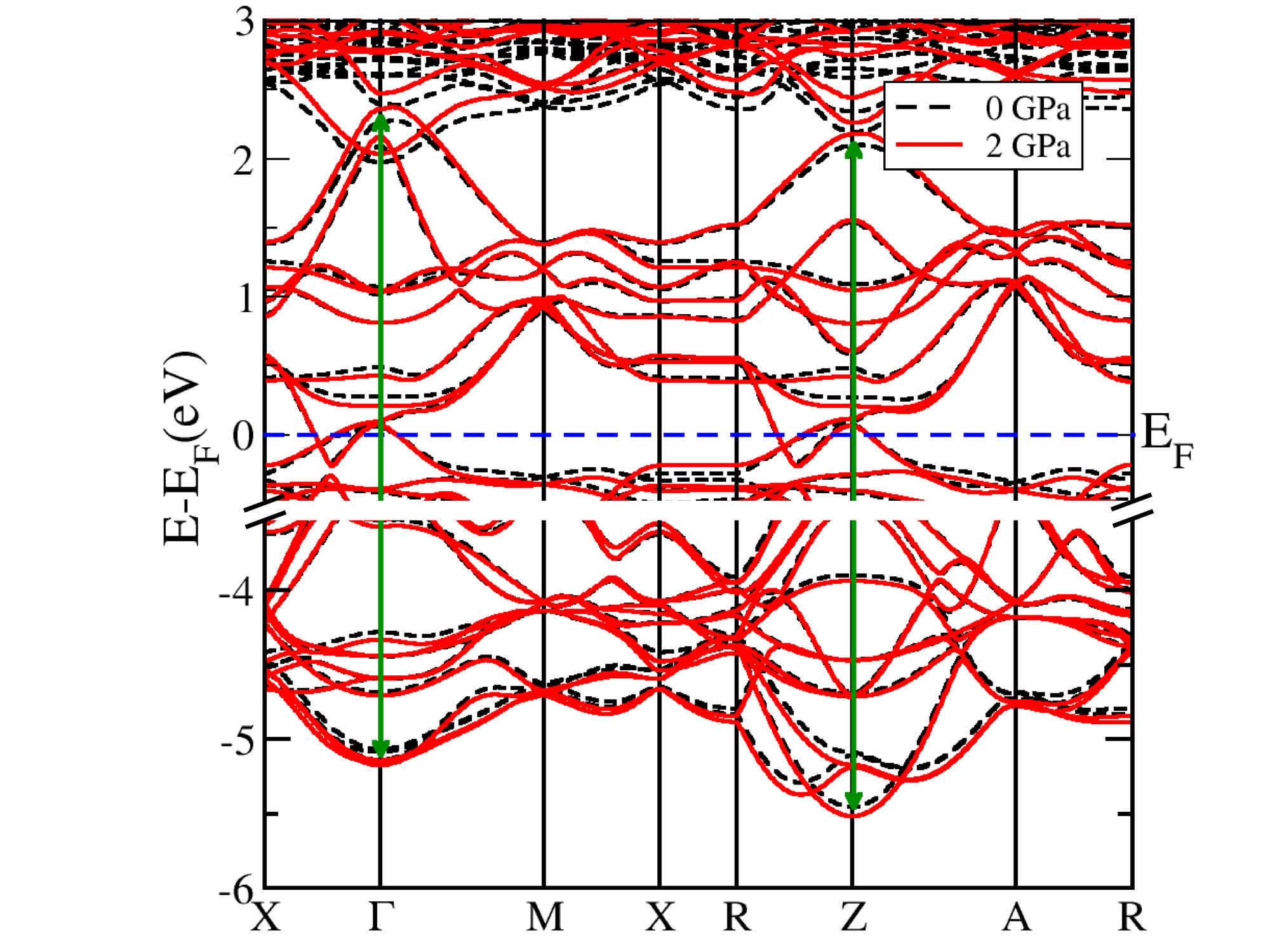}
\label{fig:band_vs_p}
\caption{(Colour on-line) LaFeAsO band structure under pressure. Continuous (red) and dashed (black) lines show the band structure for the
2 GPa compressed structure and equilibrium one respectively (Colour on-line).
}
\end{center}
\end{figure}

We are further interested in the variation of the magnetic properties with pressure. Since the experimental values for both lattice parameters and internal position are not available for all of the compounds under investigation, theoretical position are used instead, also for the ambient pressure condition. The Wien2K code allows for the optimisation of the internal degrees of freedom by the use of the Hellmann-Feymann forces. For each compound calculation of the total energy and optimised internal structure was performed for different values of the lattice parameters $a$ and $c$ for the tetragonal structure (we neglected the orthorhombic distortion).
Upon interpolation of the total energies we obtain the optimised structures at different volumes. The volume versus pressure dependence is obtained  by a fit of the Murnaghan equation of state, which yields the theoretical structural parameters.

We find that pressure reduces the magnetic moment on Fe in all three materials under investigation, confirming previous calculations for RE=Ln \cite{Opahle2009, Lebegue2009, Yang2009}. This is in agreement with the Stoner criterion: the energy gain for the magnetic state correlates inversely with the width of the $d$-bands, up to a point where the system reverts to the paramagnetic state. Indeed, in pure Fe the reduction of volume induced by applied pressure increases the width of the $d$-bands. In the present case, the maximum 2.3 GPa pressure is not sufficient to induce a paramagnetic transition, but the effect is in agreement with this qualitative argument.

This picture is confirmed in the case of LaFeAsO by the comparison of the band structures for the equilibrium structure (solid) and under 2 GPa pressure (dashed), shown in Fig.~\ref{fig:band_vs_p}. We see that the width of the valence manifold increases about the $\Gamma$ and $A$ point as evidenced by the vertical arrows.

\section{\label{sec.Disconclusions} Discussion and conclusions}

\begin{figure}[ht]
\begin{center}
\includegraphics[width=90mm]{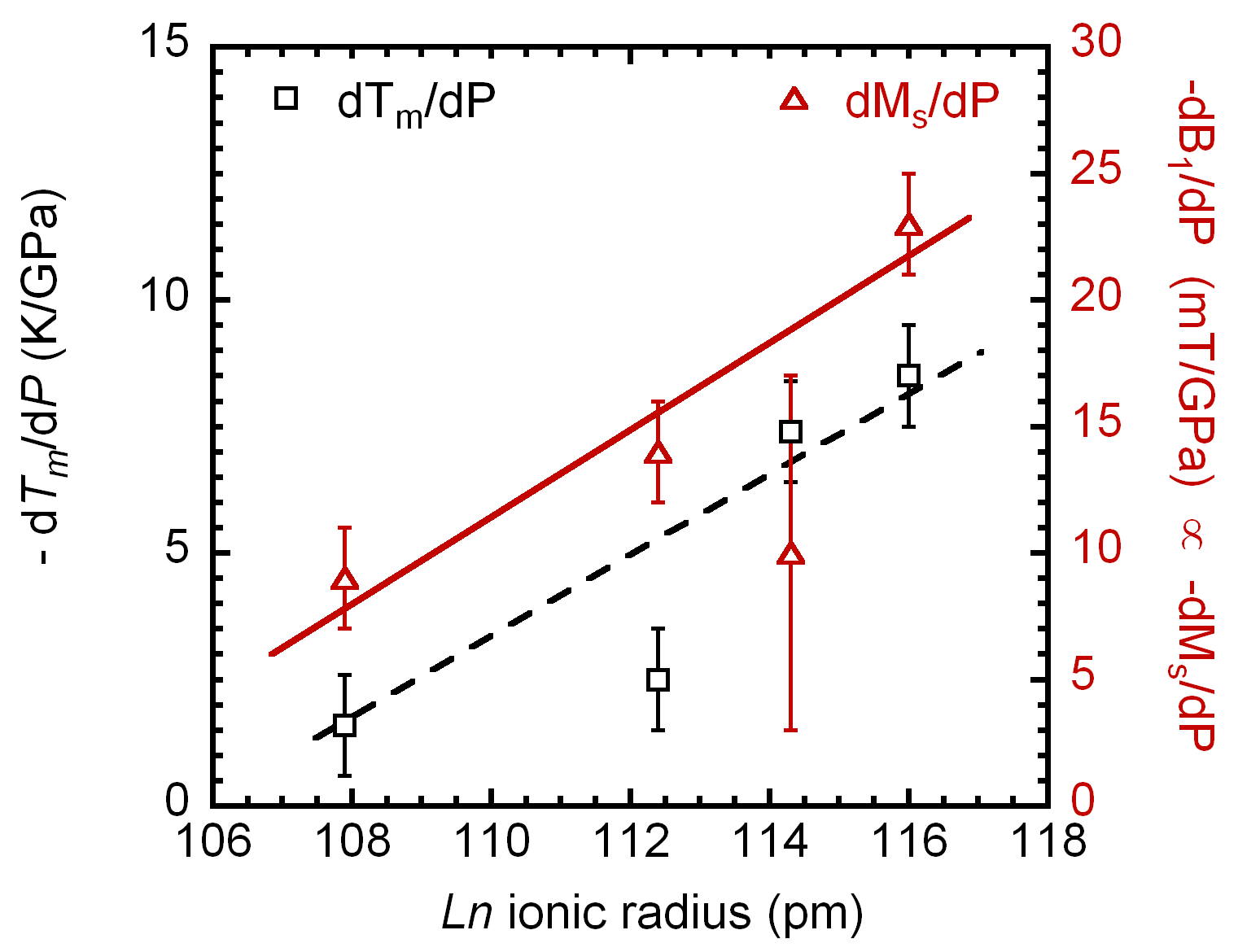}

\caption{\label{fig:P_rates}(Colour on-line) Pressure dependence of magnetic temperature of the FeAs layers, $dT_m/dp$, and staggered magnetisation at zero temperature, $dM_s/dp$, rates. Lines are guides to the eye.
}
\end{center}
\end{figure}

The ab-initio muon site identification agrees with previous estimates \cite{Maeter2009}, yielding the same determination of the iron moment, compatible with the upper end of the neutron diffraction values \cite{Lumsden2010}. Since muons conclusively determine that all Ln 1111 compounds have very close internal fields our calculations guarantee that they must have correspondingly close values of the iron local moment.

Our determination of the iron moment is obtained by comparing the calculated distant dipole field with the experimental value from $\mu$SR. In principle two additional contributions are present: the contact Fermi field, written as $\bm{B}_{F}=2\mu_0\bm{m}_{Fe}(\rho_\uparrow(0)-\rho_\downarrow(0))/3$  for collinear magnetic structures, where $\rho_{\uparrow\downarrow}(\bm{r})$ is the up(down) spin density at a position $\bm r$ and the muon site is at $r=0$; the anisotropic on-site pseudo-dipolar field, i.e. the the dipolar field averaged over the central muffin tin. We notice, however, that the three sites listed in Tab.~\ref{tab:muon} share high $x,y$ symmetry. Thanks to this condition the experimental staggered collinear structures yield $B_F=0$ by symmetry in all of them, fully justifying our neglect of this contribution. Furthermore the DFT calculations shows that the $p$-wave contribution at the muon, dominant in the pseudo-dipolar field, is also negligible. Consequently the distant dipole field is a very good estimate of $B_1$ and within this approximation $\mu$SR yields a direct measurement of the moment on iron. We note that its value, $m_{Fe} = 0.68(2)\, \mu_B$, is in very close agreement with the most recent neutron scattering \cite{Qureshi2010}, $m_{Fe} =  0.63(1)\, \mu_B$, and NMR \cite{Graf2009}, $m_{Fe} = 0.58(9)\, \mu_B$.

A more precise calculation, beyond the scope of the present work, involves the expectation value of the total local field over the muon ground state, roughly represented in Fig.~\ref{fig:muon} by the shaded volumes around each site. This is a second order effect that is certainly more significant at site B, in view of it larger and non isotropic zero point volume, compared to that of site A.
As we pointed out the dipolar field at B site (Tab.~\ref{tab:muon}) overestimates $B_2(0)$ by a factor 1.6. However, the average field at the boundary of the zero point volume in B is reduced by a factor 0.5, hence the accurate inclusion of the muon zero point motion will certainly improve the agreement between calculations and experiments also for this site. 

The results of our study show that for all the compounds under investigation, LnFeAsO with Ln = La, Ce, Pr and Sm, both the staggered magnetisation, $M_s$, and the magnetic transition temperature of the FeAs layers, $T_m$, are progressively diminished as a function of the external pressure. Up to the maximum applied pressure of $p\approx2.3$ GPa all the samples display only a modest variation of the magnetic properties, indicating that the complete destruction of magnetism requires many GPa. By DFT calculations we have shown that the reduction of the Fe magnetic moment as a function of pressure is mainly related to the broadening of the $d$-bands. As expected in an itinerant magnet such as LnFeAsO, this increases the kinetic energy and reduces the exchange energy for the spin alignment.

We experimentally found that the rate of the reduction of both the magnetic temperature, $dT_m/dp$, and the staggered magnetisation, $dM_s/dp$, are Ln dependent. Figure ~\ref{fig:P_rates} shows that both increase almost linearly as a function of the ionic radius, i.e. of the chemical pressure which increases from La to Sm. This behaviour indicates that the chemical pressure makes the structure of the latter more rigid, reducing the effect of the external pressure.  A similar effect has been recently reported for Ce alone \cite{Zocco2011}, where the magnetic transition is detected indirectly through the peak in resistivity at $T^*$, associated with the structural transition, precursor of $T_N$. The shift in $T^*$ from 145(5) to 124(5) K under the application of 2.2 GPa agrees well with that reported by us in Fig.~\ref{fig:TmP}: from 127 to 110 under 2.0 GPa. It is interesting to note that the the sensitivity of magnetism to pressure correlates inversely with the optimal superconducting $T_c$ of each Ln family.

Our results show that the effect of the external pressure on the N\'eel
transition temperature of the magnetic Ln sub-lattice is also Ln
dependent displaying an increase of $T_N^{Ln}$ for both Ce and Pr and
a decrease for Sm. Figure \ref{fig:TNLn} shows that the biggest
variation is observed for Ln=Ce, with a rate of $dT_N^{Ce}/dp=0.8(1)$
K/GPa, in agreement again with the transport measurements
reported in Ref.~\cite{Zocco2011}. This variation has been related to
the increase of the $f$-$d$ hybridization between the Fe and Ce
orbitals, which leads to the enhancement of the indirect exchange
coupling among the Ce ions \cite{Zocco2011}. The same effect in the Pr and
Sm cases is expected to be significantly smaller since the degree of $f$-$d$
hybridisation is much lower \cite{Nevidomskyy2009}. This can explain
the small increase observed for Ln=Pr in Fig.~\ref{fig:TNLn}b, which
leads to $dT_N^{Pr}/dp=0.22(5)$ K/GPa. On the other hand  Ln = Sm seems to undergo a first order magnetic transition, according to
Ref. ~\cite{Maeter2009}. If this is the case it should be accompanied by a
spontaneous deformation of the crystal structure with a concomitant
variation of the volume cell. The external pressure would then contrast the increase of the volume cell and hence lower the associated
$T_N^{Sm}$, which would explain the observed
negative $dT_N^{Sm}/dp=-0.3(1)$ K/GPa. The occurrence of a change of the volume cell deserves an
experimental confirmation.

In conclusion, we have measured the evolution of the magnetic properties of both the FeAs and LnO layers and performed DFT calculations for the Fe magnetic
moment and the muon sites in LnFeAsO (Ln = La, Ce, Pr, Sm) as a function of external pressure up to $\sim 2.3$ GPa. Our results show that the behavior of all the magnetic properties under pressure is Ln dependent. Regarding the FeAs layers, both the magnetic transition temperature $T_m$ and the Fe magnetic moment decrease as a function of pressure for all the Ln, but with a pressure rate which scales almost linearly with the Ln ionic size, being maximum for the larger Ln = La atoms.
On the other hand, the pressure dependence of the antiferromagnetic transition temperature of the LnO sublattice $T_N^{Ln}$ for magnetic Ln = Ce, Pr, Sm ions, do not follow the simple trend of Fe, but it seems to be sensitive to the different f-electron overlap. In addition Ln = Sm displays a negative $dT_N^{Sm}/dp$ rate which might be inferred to the first-order nature of the sublattice Sm transition.

\subsection{Acknowledgements}

We acknowledge partial financial support from the PRIN-08 2008XWLWF9 project and from FP7 226507 NMI3 Access. MB acknowledges support by the SNF.


\end{document}